\RequirePackage{fix-cm}
\documentclass[twocolumn]{svjour3}          
\smartqed  
\usepackage{amsmath,amssymb}
\usepackage{graphicx}
\usepackage{dcolumn}
\usepackage{bm}
\usepackage{euscript}
\usepackage{float}
\usepackage[colorlinks=true]{hyperref}
\usepackage{cancel}

\begin{document}

\title{Topology controlled phase coherence and quantum fluctuations in superconducting nanowires
}


\author{Alexey Radkevich         \and
        Andrew G. Semenov \and
        Andrei D. Zaikin 
}


\institute{A. Radkevich \at
              I.E. Tamm Department of Theoretical Physics, P.N. Lebedev Physical Institute, 119991 Moscow, Russia \\
           \and
           A.G. Semenov \at
              I.E. Tamm Department of Theoretical Physics, P.N. Lebedev Physical Institute, 119991 Moscow, Russia; \\
              National Research University Higher School of Economics, 101000 Moscow, Russia
               \and
           A.D. Zaikin \at
           Institute of Nanotechnology, Karlsruhe Institute of Technology (KIT), 76021 Karlsruhe, Germany;\\
           I.E. Tamm Department of Theoretical Physics, P.N. Lebedev Physical Institute, 119991 Moscow, Russia
}

\date{Received: date / Accepted: date}

\maketitle

\begin{abstract}
Superconducting properties of metallic nano-wires may strongly depend on specific experimental conditions. Here we consider a setup where superconducting phase fluctuations are restricted at one point inside the wire and equilibrium supercurrent flows along the wire segment of an arbitrary length $L$. Low temperature physics of this structure is essentially determined, on one hand, by smooth phase fluctuations and, on the other hand, by quantum phase slips. The zero temperature phase diagram is controlled by the wire cross section and consists of a truly superconducting phase and two different phases where superconductivity can be observed only at shorter length scales.  One of the latter phases exhibits more robust short-scale superconductivity whereas another one demonstrates a power-law decay of the supercurrent with increasing $L$ already at relatively short scales.
\end{abstract}

\section{\label{intro} Introduction}
Superconducting properties of ultrathin nanowires are markedly different from those of their bulk counterparts. The key reason for that lies in the presence of fluctuation effects which become particularly pronounced in quasi-one-dimensional superconducting systems and persist down to $T\rightarrow 0$ \cite{book,AGZ,LV}. 

Fluctuations in superconducting nanowires become progressively stronger as the wire cross section $s$ decreases. An important parameter controlling the strength of fluctuation effects is $g_\xi=R_q/R_\xi\propto s$, where $R_q=2\pi/e^2$ stands for the quantum resistance unit and $R_\xi$ defines the resistance of a wire segment of length equal to the superconducting coherence length $\xi$. Gaussian fluctuations of the superconducting order parameter give rise to a negative correction $\delta\Delta\sim \Delta/g_\xi$ to the value of the superconducting gap $\Delta$ \cite{GZ08}. The effects caused by strong (non-gaussian) fluctuations of the order parameter -- the so-called phase slips -- turn out to be even more significant. A phase slip can be viewed as localized both in space and in time full suppression of the absolute value of the order parameter accompanied by the phase jump by $\pm 2\pi$. At temperatures outside an immediate vicinity of the critical one $T_C$ such fluctuations have a quantum origin and, hence, are usually referred to as quantum phase slips (QPS). They can also be thought of as two-dimensional tunneling processes with the amplitude \cite{GZ01}
\begin{equation}
\gamma_{QPS} \sim (g_\xi\Delta/\xi)\exp (-ag_\xi), \quad a \sim 1.
\label{gamma}
\end{equation}
Compelling experimental evidence for the presence of QPS effects in superconducting nanowires was provided in a number of works \cite{BT,Lau,Zgi08}.

Another dimensionless parameter $g=R_q/Z_{\rm w}\propto \sqrt{s}$ specifically accounts for the effects associated with long range fluctuations of the superconducting phase $\varphi$. Here $Z_{\rm w}=\sqrt{\mathcal{L}_{\rm kin}/C_{\rm w}}$ is 
an effective impedance of the superconducting wire which can be viewed as a transmission line with (kinetic) inductance
$\mathcal{L}_{\rm kin}$ and (geometric) capacitance $C_{\rm w}$.  Note that smooth phase fluctuations are associated with sound-like plasma modes propagating along the wire, the so-called Mooij-Sch\"on modes \cite{MS}. Among the phenomena caused by such fluctuations is the effect of smearing of the electron density of states (DOS) \cite{RSZ17} which was recently verified experimentally \cite{Kostya}. Being controlled by $g$, this smearing can occur even at $g_\xi \gg 1$ implying both the presence of subgap electron states at any finite temperature and an effective suppression of the gap edge singularity down to its complete elimination at $g=2$ even at $T=0$.

The magnitude of logarithmic inter-QPS interactions is also controlled by the parameter $g$. For $g>16$ this interaction remains strong and, hence, ''positive'' and ''negative'' QPS remain bound in pairs. In this regime the phase coherence inside the wire is essentially preserved and the wire exhibits vanishing linear resistance. Accordingly, this state can be considered superconducting. For $g<16$, on the contrary, the inter-QPS interaction gets weaker and unbound QPS appear as a result
of Berezinskii-Kosterlitz-Thouless-like (BKT) quantum phase transition (QPT) at $g=16$. In this case the wire acquires a non-zero resistance \cite{ZGOZ}, thus signaling the absence of a superconducting response. Note, however, that such a behavior, can be observed in a specific type of experiments while in different setups the wire can exhibit superconducting properties even at $g<16$ \cite{SZ13}. A somewhat similar situation was earlier discussed in details in the case of one-dimensional arrays of resistively shunted Josephson junctions \cite{BFSZ}.

Yet another type of experiment was recently considered in \cite{RSZ19}. The corresponding setup enables passing equilibrium supercurrent across an arbitrary segment of the wire without restricting fluctuations of its superconducting phase.
It was demonstrated that the behavior of a superconducting nanowire with $g<16$ is determined by the bath of collective plasma modes and turns out to be richer than that identified earlier for different experimental setups \cite{ZGOZ,SZ13}.
In particular, for the configuration analyzed in \cite{RSZ19} total suppression of the supercurrent occurs only for $g<2$, while at $2<g<16$ the wire exhibits a mixed behavior with two different correlation lengths. This phase turns out to be superconducting at shorter length scales and non-superconducting at longer ones. 

In the present work we will further investigate equilibrium properties of a ''disordered'' phase with $g<16$  and consider the setup that allows to effectively restrict the space available for phase fluctuations  by ''pinning'' the superconducting phase $\varphi$ at one point inside the wire. We will demonstrate that such topology controlled phase pinning severely enhances the ability of the wire to conduct supercurrent. This effect can be interpreted in terms of the absence of a massless mode responsible for the destruction of superconductivity at $g<2$ in the setup considered in \cite{RSZ19}. Instead, the nanowire embedded in our present setup exhibits a transition between ''more'' and ''less'' superconducting phases characterized by different types of long-range behavior.

The structure of the paper is as follows. In Sec. \ref{model} we describe the system under consideration and present our theoretical approach. In Sec. \ref{current} we analyze the influence of smooth phase fluctuations on the supercurrent. Sec.\ref{QPS} is devoted to the effects caused by QPS. Finally, in Sec. \ref{discussion} we discuss our findings and compare them to previous results.

\section{\label{model} Model and formalism}
\begin{figure}
\includegraphics[width=\linewidth]{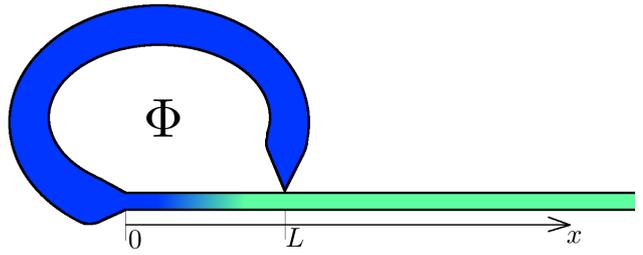}
\caption{The system under consideration}
\label{FIG1}
\end{figure}

In what follows we will restrict our attention to the setup displayed in Fig. \ref{FIG1}. It features a long and sufficiently thin superconducting nanowire with a bulk superconducting reservoir attached to one of its ends. This reservoir has a form of an open ring whose opposite end is attached to the wire by a small-area tunnel junction at a distance $L$ along the wire. The open ring is pierced by an external magnetic flux $\Phi$ which controls the phase difference $\phi=2\pi \Phi/\Phi_0$ between its ends. Accordingly, the phase at the left end of the wire is pinned by the reservoir and is set equal to zero, i.e. $\varphi(x=0)=0$.

We assume our system to be in thermodynamic equilibrium at $T \to 0$. Our setup allows one to investigate the ability of the wire to conduct non-vanishing supercurrent in a specific phase-biased measurement complimentary to those analyzed elsewhere \cite{ZGOZ,SZ13,RSZ19}. Our observable of interest is the electric current $I(\phi)$ flowing through the wire segment of length $L$ between the left wire end and the junction. As the supercurrent $I$ is a $2\pi$-periodic function of the phase $\phi$ we will restrict the phase $\phi$ to the interval $\phi\in (-\pi,\pi)$.

Low-energy physics of the system is conveniently described by the imaginary-time effective action \begin{equation}
S[\varphi]=S_{\rm w}[\varphi(x)]+S_J[\varphi(0),\varphi(L)]. \label{Eq.1}
\end{equation}
Here and below the imaginary time variable $\tau$ is omitted for brevity. The first term in this formula represents the wire effective action \cite{GZ01,ZGOZ,OGZB}
\begin{equation}
S_{\rm w}[\varphi(x)]=\frac{C_{\rm w}}{8e^2}\int\limits_0^{1/T} d\tau \int\limits_0^{\infty} dx \left[
\left(\frac{\partial\varphi}{\partial \tau}\right)^2+v^2\left(\frac{\partial\varphi}{\partial x}\right)^2
\right],
\label{wire_action}
\end{equation}
while the $S_J$-term accounts for the Josephson coupling energy and reads
\begin{equation}
S_{J}[\varphi(L)]=
-E_J\int\limits_0^{1/T} d\tau\bigl[\cos\bigl(\varphi(L)-\phi\bigr)\bigr].
\end{equation}
Integrating out the phase variable $\varphi(x)$ at all points along the wire except for its value at $x=L$ we arrive at the reduced effective action which depends only on the phase $\varphi(L,\tau)\equiv \varphi$:
\begin{equation}
S_R+S_J=\frac{1}{2}{\rm Sp} \Bigl(\varphi\, G_0^{-1}\,\varphi\Bigr)-E_J\int\limits_0^{1/T} d\tau \cos(\varphi-\phi)
\end{equation}
with
\begin{equation}
G_0(\omega_n)=\frac{8\pi}{g\omega_n}\tanh\frac{\omega_n L}{v}
\end{equation}
and $v=e^2g/(2\pi C_{\rm w})$ being respectively the bare Matsubara propagator of $\varphi$ and the velocity of the Mooij-Sch\"on mode. We observe that fluctuations of the phase variable are massive with $m_0=gv/8\pi L$. The absence of a massless mode in our setup (in contrast to that considered in \cite{RSZ19}) is a direct consequence of phase pinning at $x=0$ which prohibits uniform shifts of the phase inside the wire.

Following our previous work \cite{RSZ19} we are going to employ a variational technique in the form of a self-consistent harmonic approximation (SCHA). More details related to this method can be found, e.g., in Ref. \cite{klnrt}. Let us define the trial action
\begin{equation}
S_{\rm tr}=
\frac{1}{2}{\rm Sp}\, (\varphi-\psi)(G_{0}^{-1}+m)(\varphi-\psi). \label{ansatz2}
\end{equation}
The variational parameter $m$ accounts for the interaction-induced effective mass for the $\varphi$-mode and $\psi$ represents the average value of the phase difference.  Evaluating the free energy of the system as a function of these two parameters and minimizing it with respect to both $m$ and $\psi$, we arrive at the following SCHA equations 
\begin{align}
&E_J\cos(\psi-\phi){\rm e}^{-G(0)/2}-m=0,\label{eq_RG2}\\
&E_J\sin(\psi-\phi){\rm e}^{-G(0)/2}+\frac{gv}{8\pi L}\psi=0.\label{eq_mot2}
\end{align}
Here and below we define 
\begin{equation}
G(0)=T\sum\limits_{\omega_k}\left(G_{0}^{-1}(\omega_k)+m\right)^{-1},\label{GF}
\end{equation} 
where $\omega_k=\pi T(2k+1)$ is the Matsubara frequency.

From Eqs. (\ref{eq_RG2}), (\ref{eq_mot2}) we observe that within our variational approach the effect of phase fluctuations is fully accounted for by effective renormalization of $E_J$ by the factor ${\rm e}^{-G(0)/2}$. The supercurrent $I$ is then found from the equation
\begin{equation}
I=\frac{gev}{4\pi L}\psi .
\label{I}
\end{equation} 

\section{\label{current} Supercurrent in the presence of fluctuations}
Let us now make use of Eqs. (\ref{eq_RG2})-(\ref{I}) and explicitly evaluate the zero temperature supercurrent inside the wire segment of length $L$, see Fig. 1.

At $L\lesssim v/\Delta$ phase fluctuations are strongly suppressed and the system remains in the mean-field regime.  In the opposite limit of large $L$ the solution of Eq.(\ref{eq_RG2}) exhibits two qualitatively distinct regimes separated by $g^*=4$. At $g<4$ we find $|m|\ll v/L$. Therefore, the emergent mass is negligible, and the effect of fluctuations is purely gaussian. The equation of motion (\ref{eq_mot2}) is then rewritten as
\begin{equation}
E_J\sin(\psi-\phi)\left(\frac{\Delta L}{v}\right)^{-4/g}+\frac{gv}{8\pi L}\psi=0.\label{gaussian_eq_mot}
\end{equation}
In the interesting for us limit of small $E_J$ we may readily set $\frac{8\pi E_J}{g\Delta}< 1$. In this case the sine term is renormalized to zero faster than the kinetic inductance contribution $\propto L^{-1}$ and, hence, we obtain
\begin{equation}
I(\phi)= 2eE_J \left(\frac{v}{\Delta L}\right)^{4/g} \sin\phi.\label{I_small_g}
\end{equation}
This expression demonstrates that for $g<4$ phase fluctuations (i) modify the current-phase relation making it sine-like instead of the sawtooth-like (the latter is realized in the long $L$ limit mean-field regime) and (ii) yield a decrease of the supercurrent 
as compared to the standard Josephson formula $I(\phi)=2eE_J\sin\phi$ that applies in the limit $L\rightarrow 0$. In addition, we observe that in the presence of fluctuations the supercurrent (\ref{I_small_g}) decays faster with increasing $L$ than the standard mean field dependence $I \propto 1/L$.

Let us now turn to the case $g>4$. Resolving Eq. (\ref{eq_RG2}) in the limit $L\rightarrow\infty$ we obtain 
\begin{equation}
m=\left\{
\begin{matrix}
\left[E_J\cos(\psi-\phi)\left(\frac{8\pi}{\Delta g}\right)^\frac{4}{g}\right]^\frac{g}{g-4}, & \cos(\psi-\phi)>0, \\
-gv/8\pi L+o(1/L), & \cos(\psi-\phi)<0.
\end{matrix}
\right.\label{m}
\end{equation}
This solution remains valid only as long as $L$ exceeds the new length scale of our problem $L^*$ which reads 
\begin{equation}
L^*=\frac{v}{\Delta}\left(
\frac{g\Delta}{8\pi E_J}
\right)^\frac{g}{g-4}.\label{L*}
\end{equation}
This length scale separates the regime $L>L^*$ where fluctuations lead to a non-gaussian renormalization of the interaction potential from the gaussian regime $L \ll L^*$ where $|m|\ll gv/8\pi L$. As long as $v/\Delta\ll L\ll L^*$ the current is again given by Eq. (\ref{I_small_g}). 

For $g>4$ the renormalized Josephson coupling energy decreases slower than $1/L$ and at $L\sim L^*$ it becomes of the same order as the kinetic inductance contribution. At even larger distances the mass renormalization saturates to the value defined in Eq. (\ref{m}). The kinetic inductance contribution, on the contrary, decreases as $1/L$. Therefore, at $L\gg L^*$ the phase is pinned to the lowest minimum of the renormalized Josephson junction potential, i.e. we have $\psi =\phi$. In this case the current-phase relation reduces to the standard mean field form
\begin{equation}
I(\phi)=\frac{gev}{4\pi L}\phi.\label{mean_field_current}
\end{equation}

\section{\label{QPS} The effect of QPS}

The above analysis accounts only for the effect of smooth phase fluctuations and does not include quantum phase slips. Their influence is essentially identical to that already analyzed in Refs. \cite{SZ13,RSZ19} for different setups. Hence, at this point it suffices to only briefly summarize our key observations.

As we already discussed, for $g>16$ ''positive'' and ''negative'' quantum phase slips are bound in pairs thus causing no long-range phase decoherence. Hence, for such values of $g$ all results of the previous section remain applicable also in the presence of QPS. 

In contrast, at $g<16$ quantum phase slips are no longer bound in pairs. In this case yet another length scale appears in our problem. It plays the role of a decoherence length due to QPS and reads \cite{SZ13,RSZ19}
\begin{equation}
L_c\sim \xi \exp\left(\frac{8ag_\xi}{16-g}\right)\left(\frac{\xi \Delta}{v}\right)^{\frac{8}{16-g}}.\label{LQPS}
\end{equation}
As $L$ grows larger, quantum phase slips yield stronger disruption of long range phase coherence along the wire and, hence, faster decay of the supercurrent.  In particular, at $L\gg L_c$ we have \cite{SZ13}
\begin{equation}
I(\phi) \sim    \frac{eg_\xi\Delta \sqrt{L}}{\sqrt{\xi}}\left(\frac{v}{L \Delta}\right)^{\frac{3g}{32}} \exp\left(-\frac{3ag_\xi}{4}-\frac{L}{L_c}\right)\sin\phi .
\label{expsup}
\end{equation}
Combining this result with those derived above in the previous section we arrive at the following physical picture.

For $g<4$ there exists only one correlation length (\ref{LQPS}) in our problem. At $L\ll L_c$ QPS effects are irrelevant and the 
supercurrent suppression is merely due to smooth phase fluctuations. In this limit Eq. (\ref{I_small_g}) applies and the supercurrent decays as a power-law with increasing $L$. As soon as $L$ exceeds $L_c$ quantum phase slips come into play and the supercurrent decay becomes exponential with $L$, as it is seen in Eq. (\ref{expsup}). Thus, in practical terms the wire loses its ability to carry supercurrent in the limit of large  $L\gg L_c$. 

The situation becomes somewhat more complicated for $4<g<16$, since in this case there exist two different correlation lengths
in our problem,  $L^*$ and $L_c$. The first one diverges as $g\rightarrow 4$ while the second one tends to infinity at $g\rightarrow 16$. Depending on the relation between these two lengths, a number of different regimes can occur. 

Let us first consider the limit $L^* \ll L_c$ which can always be realized for sufficiently large values of $g_\xi$. As before, at shorter length scales $L<L^*$ only smooth phase fluctuations affect the supercurrent causing its power-law suppression with increasing $L$ and the sinusoidal current-phase relation, see Eq. (\ref{I_small_g}). At $L^* < L <L_c$ both smooth phase fluctuations and quantum phase slips are practically irrelevant and the supercurrent is defined by the standard mean field result $I \propto 1/L$ (\ref{mean_field_current}) describing the sawtooth-shaped current-phase relation. Finally, at $L \gg L_c$ the current is exponentially suppressed and the current-phase relation again reduces to the sine form, see Eq.(\ref{expsup}). 

For certain values of the system parameters -- in particular for $g$ close to $4$ -- one can also realize the regime $L^*>L_c$.
In this case there exists no room for the mean field result (\ref{mean_field_current}), whereas both Eqs. (\ref{I_small_g}) and (\ref{expsup}) remain applicable in the corresponding limits.

\section{\label{discussion} Discussion}
Superconducting properties of metallic nanowires depend not only on their parameters, but also on the topology of the experimental setup and on the way the experiment is being performed. The ability of the wire to carry supercurrent also varies at different length scales being affected by different kinds of fluctuations. In this work we investigated how fluctuations affect supercurrent in a long superconducting nanowire as a part of the setup displayed in Fig. \ref{FIG1}. This setup allows the wire to pass an equilibrium supercurrent across a segment of arbitrary length $L$ driven by an external magnetic flux. Configurations of phase inside the wire are restricted only by a bulk reservoir which pins the superconducting phase at one of the wire ends.

Fluctuation effects manifest themselves, on one hand, via sound-like collective plasma modes forming a quantum dissipative environment for electrons inside the wire and, on the other hand, via quantum phase slips. The effect of a dissipative environment formed by plasma modes boils down to an effective renormalization (reduction) of the Josephson current. As the bath of collective modes is almost Ohmic (having a lower cut-off frequency which scales as $1/L$) the low temperature system behavior resembles that involving the so-called Schmid dissipative QPT \cite{SZ90}. For $g$ smaller than the critical value ($g<4$) the effect of smooth phase fluctuations is purely gaussian. It leads to a power-law decay of the current (see Eq. (\ref{I_small_g})). The current-phase relation is sine-like in this regime. For $g>4$ a new length scale $L^*$ defined in Eq.(\ref{L*}) appears beyond which the bath obtains a finite interaction-induced mass as in Eq. (\ref{m}) and the phase becomes pinned to the value defined by an external magnetic flux. At this scale the current becomes insensitive to smooth phase fluctuations and is given by a simple mean field formula (\ref{mean_field_current}) with a sawtooth shape of current-phase dependence.

The presence of quantum phase slips naturally leads to a BKT-type quantum phase transition at $g=16$ \cite{ZGOZ}. For $g>16$ QPS do not have any significant impact on the supercurrent and the wire retains superconductivity at any length scale (with supercurrent somewhat reduced by fluctuations at $L<L^*$). At $g<16$ phase slips are unbound and cause an exponential decay of the current at $L\gg L_c$ according to Eq. (\ref{expsup}) where the phase relaxation length $L_c$ is defined in Eq. (\ref{LQPS}). Smooth phase fluctuations are irrelevant in this regime and the wire loses superconductivity at soon as $L$ strongly exceeds $L_c$. On the contrary, at $L<L_c$ QPS do not play a significant role and the physics is completely determined by smooth phase fluctuations. 

It is interesting to compare our present results with those derived recently in Ref. \cite{RSZ19}. While the setup \cite{RSZ19} allows for unrestricted fluctuations of the superconducting phase, here we consider a different topology which effectively pins 
the phase at one of the wire ends. In the former case a gapless Ohmic mode  (associated with uniform phase shifts along the wire) appears at any $L$, in contrast to the situation considered here. Fluctuations associated with this gapless mode cause a Schmid-like QPT at $g=2$. As a result, in the setup considered in Ref.\cite{RSZ19} the wire completely loses superconductivity at $g<2$, whereas the phase with $2<g<16$ is mixed, i.e. it is non-superconducting in the long length limit and superconducting at shorter scales, even though the gapless mode causes additional suppression of current in the limit $L\rightarrow 0$. Comparing this situation with the one considered here, we observe that superconductivity is severely enhanced by the phase pinning as a result of the soft mode suppression. This effect turns the QPT at $g=2$  \cite{RSZ19} into a transition between ''less'' and ''more'' superconducting phases at $g=4$ considered here. We also note that a similar phase transition was also discussed in Ref. \cite{HG} in the context of superconducting nanorings interrupted by a Josephson junction.

It would be interesting to verify our predictions in experiments with superconducting nanowires.

ADZ and AGS acknowledge the financial support by RFBR Grant No. 18-02-00586. 
AR acknowledges the RFBR Grant No. 19-32-90229.

\end{document}